# Quasi-Orthogonal STBC with Minimum Decoding Complexity

Chau Yuen, Yong Liang Guan, Tjeng Thiang Tjhung

*Abstract* — In this paper, we consider a Quasi-Orthogonal STBC with minimum decoding complexity (MDC-QOSTBC). We formulate its algebraic structure and propose a systematic method for its construction. We show that a maximum likelihood (ML) decoder for this MDC-QOSTBC for any numbers of transmit antennas only requires the joint detection of two real symbols. Assuming the use of a square or rectangular QAM or MPSK modulation for this MDC-QOSTBC, we also obtain the optimum constellation rotation angle in order to achieve full diversity and optimum coding gain. We show that the maximum achievable code rate of these MDC-QOSTBC is 1 for three and four antennas, and ¾ for five to eight antennas. We also show that the proposed MDC-QOSTBC has several desirable properties, such as more even power distribution among antennas and better scalability in adjusting the number of transmit antennas compared with the Co-ordinate Interleaved Orthogonal Design (CIOD) and Asymmetric CIOD codes. For the case of an odd number of transmit antennas, MDC-QOSTBC also has better decoding performance than CIOD.

I. INTRODUCTION

Orthogonal Space-Time Block Code (O-STBC) that can offer full transmit diversity and linear decoding complexity has been designed in [1,2,12]. Unfortunately, O-STBCs suffer from a reduced code rate when complex signal constellations and more than two transmit antennas are used [1,2,12]. Therefore STBC designs that can achieve full transmit diversity and higher code rate but requiring only moderate decoding complexity are desirable.

To this end, some Quasi-Orthogonal STBC (QO-STBC) with constellation rotation has been proposed in [3-6] that is able to achieve full code rate by relaxing the strict orthogonality requirement of O-STBC. The maximum-likelihood (ML) decoding of QO-STBC can be performed by searching over only pairs, instead of the full set, of the possible transmitted complex symbols. Subsequently, Coordinate Interleaved Orthogonal Design (CIOD) and Asymmetric CIOD (ACIOD) have been proposed in [7,8] to provide high code rate and full transmit diversity (after constellation rotation) with even lower decoding complexity. However, these codes require up to half of the transmit antennas to be turned off regularly, thus introducing high peak-to-average transmitter power ratio which is undesirable [8,11].

In this paper, we focus on a new class of QO-STBC whose ML decoding only requires the joint detection of two *real* symbols. This is the lowest possible decoding complexity for any non-orthogonal STBCs. Hence we call it Minimum-Decoding-Complexity QO-STBC (MDC-QOSTBC). We shall derive its algebraic structure, propose systematic methods to construct it, and investigate its maximum achievable code rate. We will also compare its decoding performance, power distribution properties (which is related to the number of antennas to be turned off regularly) and antenna scalability (scalability in supporting different number of transmit antennas) with the existing QO-STBCs, CIOD and ACIOD.

## II. SIGNAL MODEL AND O-STBC

### A. Generic STBC

Suppose that a generic STBC codeword is transmitted from $N_t$ transmit antennas to $N_r$ receive antennas over an interval of $T$ symbol periods in which the propagation channel condition is time-invariant and known to the receiver. The transmitted codeword can be written as a $T \times N_t$ matrix **G** that consists of $K$ arbitrary complex constellation symbols. Its code rate is defined as

$R = K/T$. Following the model in [9], **G** can be expressed as:

$$\mathbf{G} = \sum_{q=1}^{K}(x_q^R \mathbf{A}_q + jx_q^I \mathbf{B}_q) \quad (1)$$

where the transmitted symbols are $x_q = x_q^R + jx_q^I$, and the superscripts $(\ )^R$ and $(\ )^I$ denote respectively, the real and imaginary part of a complex element, vector or matrix. Matrices $\mathbf{A}_q$ and $\mathbf{B}_q$ are called the "dispersion matrices" and are of size $T \times N_t$. For the given numbers of transmit antennas, the design of a STBC depends crucially on the choices of the parameters $T$, $K$, and the dispersion matrices $\{\mathbf{A}_q, \mathbf{B}_q\}$. The transmitted and received signals are related by [9]:

$$\tilde{\mathbf{r}} = \sqrt{\rho/N_t}\mathbf{H}\tilde{\mathbf{x}} + \tilde{\boldsymbol{\eta}} \quad (2)$$

where $\tilde{\mathbf{r}} = \begin{bmatrix} \mathbf{r}_1^R \\ \mathbf{r}_1^I \\ \vdots \\ \mathbf{r}_{N_r}^R \\ \mathbf{r}_{N_r}^I \end{bmatrix}, \tilde{\mathbf{x}} = \begin{bmatrix} x_1^R \\ x_1^I \\ \vdots \\ x_K^R \\ x_K^I \end{bmatrix}, \tilde{\boldsymbol{\eta}} = \begin{bmatrix} \boldsymbol{\eta}_1^R \\ \boldsymbol{\eta}_1^I \\ \vdots \\ \boldsymbol{\eta}_{N_r}^R \\ \boldsymbol{\eta}_{N_r}^I \end{bmatrix}, \mathbf{H} = \begin{bmatrix} \mathcal{A}_1\underline{\mathbf{h}}_1 & \mathcal{B}_1\underline{\mathbf{h}}_1 & \cdots & \mathcal{A}_K\underline{\mathbf{h}}_1 & \mathcal{B}_K\underline{\mathbf{h}}_1 \\ \vdots & \vdots & \ddots & \vdots & \vdots \\ \mathcal{A}_1\underline{\mathbf{h}}_{N_r} & \mathcal{B}_1\underline{\mathbf{h}}_{N_r} & \cdots & \mathcal{A}_K\underline{\mathbf{h}}_{N_r} & \mathcal{B}_K\underline{\mathbf{h}}_{N_r} \end{bmatrix},$

$\mathcal{A}_q = \begin{bmatrix} \mathbf{A}_q^R & -\mathbf{A}_q^I \\ \mathbf{A}_q^I & \mathbf{A}_q^R \end{bmatrix}, \mathcal{B}_q = \begin{bmatrix} -\mathbf{B}_q^I & -\mathbf{B}_q^R \\ \mathbf{B}_q^R & -\mathbf{B}_q^I \end{bmatrix}, \underline{\mathbf{h}}_i = \begin{bmatrix} \mathbf{h}_i^R \\ \mathbf{h}_i^I \end{bmatrix}.$

In the above equation, $\mathbf{r}_i$ and $\boldsymbol{\eta}_i$ ($1 \leq i \leq N_r$) are $T \times 1$ column vectors which contain the received signal and AWGN noise for the $i^{th}$ receive antenna respectively, over $T$ symbol periods. **H** is called the equivalent channel matrix, $\mathbf{h}_i$ is $N_t \times 1$ column vector that contains the fading coefficients of the spatial sub-channels between the $N_t$ transmit antennas and $i^{th}$ receive antenna. The normalization factor $\sqrt{\rho/N_t}$ in (2) ensures that $\rho$ is the signal-to-noise ratio (SNR) at each receive antenna, regardless of whatever $N_t$ is.

### B.  Orthogonal STBC

Orthogonal STBC (O-STBC) has the simplest decoding complexity, as its ML decoding can be achieved by linear detection. It has been shown in [2] that to design an O-STBC is equivalent to finding $K$ sets of dispersion matrices $\{\underline{\mathbf{A}}_q, \underline{\mathbf{B}}_q\}$ (in this paper the underlined dispersion matrices are meant for an O-STBC, while the dispersion matrices of the MDC-STBC's are not underlined), which satisfy:

$$\begin{align}&\text{(i)} \ \underline{\mathbf{A}}_q^H \underline{\mathbf{A}}_q = \mathbf{I}_{N_t} \quad , \ \underline{\mathbf{B}}_q^H \underline{\mathbf{B}}_q = \mathbf{I}_{N_t} \quad && 1 \leq q \leq K \\ &\text{(ii)} \ \underline{\mathbf{A}}_q^H \underline{\mathbf{A}}_p = -\underline{\mathbf{A}}_p^H \underline{\mathbf{A}}_q \ , \ \underline{\mathbf{B}}_q^H \underline{\mathbf{B}}_p = -\underline{\mathbf{B}}_p^H \underline{\mathbf{B}}_q \quad && 1 \leq q \neq p \leq K \\ &\text{(iii)} \ \underline{\mathbf{A}}_q^H \underline{\mathbf{B}}_p = \underline{\mathbf{B}}_p^H \underline{\mathbf{A}}_q \quad && 1 \leq q, p \leq K \end{align} \quad (3)$$

### III. ALGEBRAIC STRUCTURE OF MDC-QOSTBC

In MDC-QOSTBC, the goal is to divide the transmitted symbols into $K$ independent groups ($K$ = number of complex symbols transmitted in one block), such that every complex symbol is orthogonal to all other complex symbols, but the I and Q components within the same complex symbol need not be orthogonal. As a result of such grouping, the received symbols can be separated into $K$ independent groups by simple linear processing or matched filtering, ML decoding of different groups can then be performed separately, and in parallel. In each group, only two real symbols (i.e. the I and Q components) need to be jointly detected.

*Definition 1*: A *Minimum-Decoding-Complexity QO-STBC* (MDC-QOSTBC) is a QO-STBC such that its equivalent channel matrix $\mathbf{H}$ has the property that $\mathbf{H}^T\mathbf{H}$ is block-diagonal with non-zero sub-matrices of size 2×2.

It should be noted from *Definition 1* that the $\mathbf{H}^T\mathbf{H}$ of an O-STBC [1,2,12] is a diagonal matrix, while the $\mathbf{H}^T\mathbf{H}$ of a QO-STBC which needs the joint detection of $s$ real symbols will be block-diagonal with $s \times s$ sub-block matrices. Therefore, MDC-QOSTBC has the minimum decoding complexity among all non-orthogonal STBC, because it only needs the joint detection of two real symbols, anything less complex (i.e. linear detection of only one real symbol) would be an O-STBC.

Next we derive the algebraic structure of MDC-QOSTBC. At the receiver, a matched filter $\mathbf{H}^T$ is multiplied to the received signal $\tilde{\mathbf{r}}$ in (2) to separate the received symbols into $K$ independent groups. Let us consider a snapshot of $\mathbf{H}^T\mathbf{H}$ as follows:

$$\mathbf{H}^{\mathrm{T}}\mathbf{H} = \begin{bmatrix} \vdots & \cdots & \vdots \\ \underline{\mathbf{h}}_1^{\mathrm{T}}\mathcal{A}_q^{\mathrm{T}} & \cdots & \underline{\mathbf{h}}_{N_r}^{\mathrm{T}}\mathcal{A}_q^{\mathrm{T}} \\ \underline{\mathbf{h}}_1^{\mathrm{T}}\mathcal{B}_q^{\mathrm{T}} & \cdots & \underline{\mathbf{h}}_{N_r}^{\mathrm{T}}\mathcal{B}_q^{\mathrm{T}} \\ \underline{\mathbf{h}}_1^{\mathrm{T}}\mathcal{A}_p^{\mathrm{T}} & \cdots & \underline{\mathbf{h}}_{N_r}^{\mathrm{T}}\mathcal{A}_p^{\mathrm{T}} \\ \underline{\mathbf{h}}_1^{\mathrm{T}}\mathcal{B}_p^{\mathrm{T}} & \cdots & \underline{\mathbf{h}}_{N_r}^{\mathrm{T}}\mathcal{B}_p^{\mathrm{T}} \\ \vdots & \cdots & \vdots \end{bmatrix} \begin{bmatrix} \cdots & \mathcal{A}_q\underline{\mathbf{h}}_1 & \mathcal{B}_q\underline{\mathbf{h}}_1 & \mathcal{A}_p\underline{\mathbf{h}}_1 & \mathcal{B}_p\underline{\mathbf{h}}_1 & \cdots \\ \vdots & \vdots & \vdots & \vdots & \vdots & \vdots \\ \cdots & \mathcal{A}_q\underline{\mathbf{h}}_{N_r} & \mathcal{B}_q\underline{\mathbf{h}}_{N_r} & \mathcal{A}_p\underline{\mathbf{h}}_{N_r} & \mathcal{B}_p\underline{\mathbf{h}}_{N_r} & \cdots \end{bmatrix} \qquad 1 \leq q \neq p \leq K$$

$$= \begin{bmatrix} \cdots & \cdots & \cdots & \cdots & \cdots & \cdots \\ \vdots & \sum_{i=1}^{N_R}\underline{\mathbf{h}}_i^{\mathrm{T}}(\mathcal{A}_q^{\mathrm{T}}\mathcal{A}_q)\underline{\mathbf{h}}_i & \sum_{i=1}^{N_R}\underline{\mathbf{h}}_i^{\mathrm{T}}(\mathcal{A}_q^{\mathrm{T}}\mathcal{B}_q)\underline{\mathbf{h}}_i & \boxed{\sum_{i=1}^{N_R}\underline{\mathbf{h}}_i^{\mathrm{T}}(\mathcal{A}_q^{\mathrm{T}}\mathcal{A}_p)\underline{\mathbf{h}}_i \quad \sum_{i=1}^{N_R}\underline{\mathbf{h}}_i^{\mathrm{T}}(\mathcal{A}_q^{\mathrm{T}}\mathcal{B}_p)\underline{\mathbf{h}}_i} & \vdots \\ \vdots & \sum_{i=1}^{N_R}\underline{\mathbf{h}}_i^{\mathrm{T}}(\mathcal{B}_q^{\mathrm{T}}\mathcal{A}_q)\underline{\mathbf{h}}_i & \sum_{i=1}^{N_R}\underline{\mathbf{h}}_i^{\mathrm{T}}(\mathcal{B}_q^{\mathrm{T}}\mathcal{B}_q)\underline{\mathbf{h}}_i & \boxed{\sum_{i=1}^{N_R}\underline{\mathbf{h}}_i^{\mathrm{T}}(\mathcal{B}_q^{\mathrm{T}}\mathcal{A}_p)\underline{\mathbf{h}}_i \quad \sum_{i=1}^{N_R}\underline{\mathbf{h}}_i^{\mathrm{T}}(\mathcal{B}_q^{\mathrm{T}}\mathcal{B}_p)\underline{\mathbf{h}}_i} & \vdots \\ \vdots & \boxed{\sum_{i=1}^{N_R}\underline{\mathbf{h}}_i^{\mathrm{T}}(\mathcal{A}_p^{\mathrm{T}}\mathcal{A}_q)\underline{\mathbf{h}}_i \quad \sum_{i=1}^{N_R}\underline{\mathbf{h}}_i^{\mathrm{T}}(\mathcal{A}_p^{\mathrm{T}}\mathcal{B}_q)\underline{\mathbf{h}}_i} & \sum_{i=1}^{N_R}\underline{\mathbf{h}}_i^{\mathrm{T}}(\mathcal{A}_p^{\mathrm{T}}\mathcal{A}_p)\underline{\mathbf{h}}_i & \sum_{i=1}^{N_R}\underline{\mathbf{h}}_i^{\mathrm{T}}(\mathcal{A}_p^{\mathrm{T}}\mathcal{B}_p)\underline{\mathbf{h}}_i & \vdots \\ \vdots & \boxed{\sum_{i=1}^{N_R}\underline{\mathbf{h}}_i^{\mathrm{T}}(\mathcal{B}_p^{\mathrm{T}}\mathcal{A}_q)\underline{\mathbf{h}}_i \quad \sum_{i=1}^{N_R}\underline{\mathbf{h}}_i^{\mathrm{T}}(\mathcal{B}_p^{\mathrm{T}}\mathcal{B}_q)\underline{\mathbf{h}}_i} & \sum_{i=1}^{N_R}\underline{\mathbf{h}}_i^{\mathrm{T}}(\mathcal{B}_p^{\mathrm{T}}\mathcal{A}_p)\underline{\mathbf{h}}_i & \sum_{i=1}^{N_R}\underline{\mathbf{h}}_i^{\mathrm{T}}(\mathcal{B}_p^{\mathrm{T}}\mathcal{B}_p)\underline{\mathbf{h}}_i & \vdots \\ \cdots & \cdots & \cdots & \cdots & \cdots & \cdots \end{bmatrix} \quad (4)$$

To comply with Definition 1, the boxed summation terms in (4) must all be zero. To achieve this, $\mathcal{A}_q^{\mathrm{T}}\mathcal{A}_p$, $\mathcal{A}_q^{\mathrm{T}}\mathcal{B}_p$, $\mathcal{B}_q^{\mathrm{T}}\mathcal{A}_p$, $\mathcal{B}_q^{\mathrm{T}}\mathcal{B}_p$ must be skew-symmetric, as a result of *Theorem 1* stated below.

<u>*Theorem 1*</u>: For any vector $\mathbf{v}$, $\mathbf{v}^{\mathrm{T}}\mathbf{M}\mathbf{v} = 0$ if the matrix $\mathbf{M}$ is skew-symmetric, i.e. $\mathbf{M}^{\mathrm{T}} = -\mathbf{M}$.

<u>Proof of *Theorem 1*</u>: Let $\mathbf{v}^{\mathrm{T}}\mathbf{M}\mathbf{v} = c$. Since $c$ is a constant value, $c^{\mathrm{T}} = c = \mathbf{v}^{\mathrm{T}}\mathbf{M}^{\mathrm{T}}\mathbf{v} = -\mathbf{v}^{\mathrm{T}}\mathbf{M}\mathbf{v}$ (if $\mathbf{M}^{\mathrm{T}} = -\mathbf{M}$). Now, $c + c^{\mathrm{T}} = \mathbf{v}^{\mathrm{T}}\mathbf{M}\mathbf{v} - \mathbf{v}^{\mathrm{T}}\mathbf{M}\mathbf{v} = 0$, so $c = 0$, hence *Theorem 1* is proved. ∎

<u>*Theorem 2*</u>: For different complex symbols (indexed using subscripts $q$ and $p$) in an MDC-STBC to be orthogonal to each other, i.e. $\mathcal{A}_q^{\mathrm{T}}\mathcal{A}_p$, $\mathcal{A}_q^{\mathrm{T}}\mathcal{B}_p$, $\mathcal{B}_q^{\mathrm{T}}\mathcal{A}_p$, $\mathcal{B}_q^{\mathrm{T}}\mathcal{B}_p$ to be skew-symmetric, their dispersion matrices $\{\mathbf{A}_q, \mathbf{B}_q\}$ and $\{\mathbf{A}_p, \mathbf{B}_p\}$ must possess the following algebraic structure, herein referred as *Minimum Decoding Complexity Quasi-Orthogonality* (MDC-QO) *Constraints*:

$$\begin{aligned} &\text{(i)} \ \mathbf{A}_q^{\mathrm{H}}\mathbf{A}_p = -\mathbf{A}_p^{\mathrm{H}}\mathbf{A}_q, \ \mathbf{B}_q^{\mathrm{H}}\mathbf{B}_p = -\mathbf{B}_p^{\mathrm{H}}\mathbf{B}_q \\ &\text{(ii)} \ \mathbf{A}_q^{\mathrm{H}}\mathbf{B}_p = \mathbf{B}_p^{\mathrm{H}}\mathbf{A}_q \end{aligned} \qquad 1 \leq q \neq p \leq K \qquad (5)$$

<u>Proof of *Theorem 2*</u>: We take the MDC-QO Constraints (5)(i) as an example:

$$\mathbf{A}_q^H \mathbf{A}_p = -\mathbf{A}_p^H \mathbf{A}_q$$
$$\Rightarrow (\mathbf{A}_q^R + j\mathbf{A}_q^I)^H (\mathbf{A}_p^R + j\mathbf{A}_p^I) = -(\mathbf{A}_p^R + j\mathbf{A}_p^I)^H (\mathbf{A}_q^R + j\mathbf{A}_q^I)$$
$$\Rightarrow \begin{cases} \left(\mathbf{A}_q^R\right)^T \mathbf{A}_p^R + \left(\mathbf{A}_q^I\right)^T \mathbf{A}_p^I = -\left(\mathbf{A}_p^R\right)^T \mathbf{A}_q^R - \left(\mathbf{A}_p^I\right)^T \mathbf{A}_q^I \text{ (real part)} \\ \left(\mathbf{A}_q^R\right)^T \mathbf{A}_p^I - \left(\mathbf{A}_q^I\right)^T \mathbf{A}_p^R = -\left(\mathbf{A}_p^R\right)^T \mathbf{A}_q^I + \left(\mathbf{A}_p^I\right)^T \mathbf{A}_q^R \text{ (imag part)} \end{cases}$$

A skew-symmetric matrix **M** and a symmetric matrix **N** can then be defined as follows:

$$\Rightarrow \begin{cases} \mathbf{M} = \left(\mathbf{A}_q^R\right)^T \mathbf{A}_p^R + \left(\mathbf{A}_q^I\right)^T \mathbf{A}_p^I & \text{(Skew-symmetric)} \\ \mathbf{N} = \left(\mathbf{A}_q^R\right)^T \mathbf{A}_p^I - \left(\mathbf{A}_q^I\right)^T \mathbf{A}_p^R & \text{(Symmetric)} \end{cases}$$

One can show that $\mathcal{A}_q^T \mathcal{A}_p = \begin{bmatrix} \mathbf{M} & -\mathbf{N} \\ \mathbf{N} & \mathbf{M} \end{bmatrix}$, hence (5)(i) ensures that $\mathcal{A}_q^T \mathcal{A}_p$ is skew-symmetric.

Similarly $\mathcal{A}_q^T \mathcal{B}_p$, $\mathcal{B}_q^T \mathcal{A}_p$, $\mathcal{B}_q^T \mathcal{B}_p$ can be proven to be skew-symmetric as long as MDC-QO Constraints in (5) are fulfilled. Hence *Theorem 2* is proved. ∎

Note that the difference between the properties of O-STBC in (3) and MDC-QO Constraints in (5) is that (3)(iii) holds for all *k* and *p*, whereas (5)(iii) holds only when $k \neq p$. In addition, the condition (3)(i) is not required for the MDC-QO constraint because it affects the diversity order and not the decoding complexity.

It can be easily verified that all the CIOD and ACIOD codes from [7,8] comply with the algebraic structure stated in *Theorem 2*, although they were not designed from this approach. This shows that our proposed MDC-QO Constraints are generic and inclusive.

## IV.     MDC-QOSTBC

### A.     *Construction of MDC-QOSTBC from O-STBC*

In this section, we propose a systematic method to construct an MDC-QOSTBC from an O-STBC. The proposed method consists of four mapping rules, as listed in *Theorem 3* below, to map the dispersion matrices of an O-STBC to the dispersion matrices of an MDC-QOSTBC.

*Theorem 3*: Consider an O-STBC with code length $T$ for $N_t$ transmit antennas, which consists of $K$ sets of dispersion matrices denoted as $\{\underline{\mathbf{A}}_q, \underline{\mathbf{B}}_q\}$, $1 \leq q \leq K$. An MDC-QOSTBC with code length $2T$ for $2N_t$ transmit antennas, which consists of $2K$ sets of dispersion matrices denoted as $\{\mathbf{A}_q, \mathbf{B}_q\}$, where $1 \leq q \leq 2K$, can be constructed with the following four mapping rules:

$$\text{Rule 1: } \mathbf{A}_q = \begin{bmatrix} \underline{\mathbf{A}}_q & 0 \\ 0 & \underline{\mathbf{A}}_q \end{bmatrix}; \quad \text{Rule 3: } \mathbf{A}_{K+q} = \begin{bmatrix} j\underline{\mathbf{B}}_q & 0 \\ 0 & j\underline{\mathbf{B}}_q \end{bmatrix};$$

$$1 \leq q \leq K \qquad (6)$$

$$\text{Rule 2: } \mathbf{B}_q = \begin{bmatrix} 0 & j\underline{\mathbf{A}}_q \\ j\underline{\mathbf{A}}_q & 0 \end{bmatrix}; \quad \text{Rule 4: } \mathbf{B}_{K+q} = \begin{bmatrix} 0 & \underline{\mathbf{B}}_q \\ \underline{\mathbf{B}}_q & 0 \end{bmatrix}.$$

Proof of *Theorem 3*: Based on the structure of the O-STBC's dispersion matrices $\{\underline{\mathbf{A}}_q, \underline{\mathbf{B}}_q\}$ specified in (3), it can be proven that the mapping rules in (6) result in a new set of dispersion matrices $\{\mathbf{A}_q, \mathbf{B}_q\}$ that satisfy the MDC-QO Constraints in (5). Hence an MDC-QOSTBC can be constructed accordingly. The detailed proof is omitted, as the verifications are routine. ∎

A graphical example to illustrate the construction of an MDC-QOSTBC for four transmit antennas from the Alamouti O-STBC for two transmit antennas [1] is shown in Figure 1, where $\underline{\mathbf{A}}_1, \underline{\mathbf{A}}_2, \underline{\mathbf{B}}_1, \underline{\mathbf{B}}_2$ denote the dispersion matrices of the Alamouti O-STBC, while $\mathbf{A}_1$ to $\mathbf{A}_4$, $\mathbf{B}_1$ to $\mathbf{B}_4$ denote the dispersion matrices of the newly constructed MDC-QOSTBC. The codeword $\mathbf{G}$ of the resultant MDC-QOSTBC is shown in (7). It can be shown that its ML decoding metric can be calculated as the sum $f_1 + f_2 + f_3 + f_4$, where the terms $f_1$ to $f_4$ are given in (8). Since each $f_i$ is just a function of $x_i^R$ and $x_i^I$ for $1 \leq i \leq 4$ (i.e. joint detection of two real symbols), and is independent of $x_k$ for $i \neq k$, the minimization of the ML metric is equivalent to minimizing the four $f_i$ terms independently. This implies a lower decoding complexity as compared to the existing QO-STBCs [3-6].

$$G = \begin{bmatrix} x_1^R + jx_3^R & x_2^R + jx_4^R & -x_1^I + jx_3^I & -x_2^I + jx_4^I \\ -x_2^R + jx_4^R & x_1^R - jx_3^R & x_2^I + jx_4^I & -x_1^I - jx_3^I \\ -x_1^I + jx_3^I & -x_2^I + jx_4^I & x_1^R + jx_3^R & x_2^R + jx_4^R \\ x_2^I + jx_4^I & -x_1^I - jx_3^I & -x_2^R + jx_4^R & x_1^R - jx_3^R \end{bmatrix} \quad (7)$$

$$\begin{aligned} f_1(x_1) &= \sum_{r=1}^{N_r} \left[ (\sum_{n=1}^{4} |h_{n,r}|^2)(|x_1^R|^2 + |x_1^I|^2) + 2\operatorname{Re}\{x_1^R(\alpha) - x_1^I(\beta) - x_1^R x_1^I(\gamma)\} \right] \\ f_2(x_2) &= \sum_{r=1}^{N_r} \left[ (\sum_{n=1}^{4} |h_{n,r}|^2)(|x_2^R|^2 + |x_2^I|^2) + 2\operatorname{Re}\{x_2^R(\chi) - x_2^I(\delta) - x_2^R x_2^I(\varphi)\} \right] \\ f_3(x_3) &= \sum_{r=1}^{N_r} \left[ (\sum_{n=1}^{4} |h_{n,r}|^2)(|x_3^R|^2 + |x_3^I|^2) + 2\operatorname{Re}\{jx_3^R(\alpha) + jx_3^I(\beta) + x_3^R x_3^I(\gamma)\} \right] \\ f_4(x_4) &= \sum_{r=1}^{N_r} \left[ (\sum_{n=1}^{4} |h_{n,r}|^2)(|x_4^R|^2 + |x_4^I|^2) + 2\operatorname{Re}\{jx_4^R(\chi) + jx_4^I(\delta) + x_4^R x_4^I(\varphi)\} \right] \end{aligned} \quad (8)$$

where $\alpha = -h_{1,r}r_1^* - h_{2,r}^*r_2 - h_{3,r}r_3^* - h_{4,r}^*r_4$, $\beta = -h_{3,r}r_1^* - h_{4,r}^*r_2 - h_{1,r}r_3^* - h_{2,r}^*r_4$, $\gamma = 2\operatorname{Re}(h_{1,r}h_{3,r}^* + h_{2,r}h_{4,r}^*)$, $\chi = -h_{2,r}r_1^* + h_{1,r}^*r_2 - h_{4,r}r_3^* + h_{3,r}^*r_4$, $\delta = -h_{4,r}r_1^* + h_{3,r}^*r_2 - h_{2,r}r_3^* + h_{1,r}^*r_4$, $\varphi = 2\operatorname{Re}(h_{1,r}h_{3,r}^* + h_{2,r}h_{4,r}^*)$, and $h_{i,r}$ represents the fading coefficient from the $i^{th}$ transmit antenna to the $r^{th}$ receive antenna.

Similar to the QO-STBCs proposed in [3-6] and CIOD/ACIOD designs proposed in [7,8], MDC-QOSTBC constructed from *Theorem 3* cannot achieve full transmit diversity directly. We therefore use the constellation rotation technique proposed in [4-6] to attain full diversity, as well to optimize the decoding performance of the MDC-QOSTBC. The optimum angle of constellation rotation for the MDC-QOSTBC constructed by *Theorem 3* can been found analytically to be $[\tan^{-1}(1/2)]/2 = 13.29^0$ for *all* the transmit symbols of *any square or rectangular-QAM constellation* [13]. The optimum angle of rotation for QPSK and 8PSK has also been found to be $31.7^0$ and $4.9^0$ respectively [13].

### B. MDC-QOSTBC for Odd Number of Transmit Antennas

Although the construction method in *Theorem 3* specifies how to construct MDC-QOSTBC for even number of transmit antennas, we can easily prove that by removing *any* number of columns from the codeword of an MDC-QOSTBC with full diversity, the resultant code is a valid MDC-QOSTBC with full diversity that supports a smaller number of transmit antennas at the same code rate (as it fulfills the MDC-QO Constraint in (5)) [13]. For example, by removing the last column of **G** in (7), an MDC-QOSTBC for three transmit antennas is obtained.

### C.  *Maximum Achievable Code Rate of MDC-QOSTBC*

Based on *Theorem 3*, an MDC-QOSTBC for $2N_t$ transmit antennas will consist of $2K$ dispersion matrices, each of duration $2T$. Hence its code rate is $K/T$, which is the same as the code rate of the lower-order O-STBC used to generate it. Based on the maximum achievable code rate of O-STBC in [12], the maximum achievable code rate of MDC-QOSTBC can be found to be [13]:

$$R_{\text{MDC-QOSTBC}} = \frac{1+n}{2n} \quad \text{where } n = \left\lceil \frac{N_t}{4} \right\rceil \tag{9}$$

where $\lceil x \rceil$ denotes the smallest integer larger than x.

As a result, the MDC-QOSTBC for four transmit antennas (and its variant for three antennas) specified in (7) has a maximum achievable code rate of 1 (same as O-STBC for two transmit antennas [1]), while MDC-QOSTBC for eight transmit antennas (and its variants for five to seven antennas), has a maximum achievable code rate of ¾ (same as O-STBC for four transmit antennas [1,2]).

In Table 1, we give a comparison of the maximum achievable code rate and decoding complexity (i.e. the number of real symbols required for joint ML detection) of MDC-QOSTBC versus the O-STBC, QO-STBC and CIOD/ACIOD with constellation rotation. The comparison shows that our proposed MDC-QOSTBC achieves:

1) higher code rate than O-STBC with the same diversity level (number of transmit antennas);

2) lower decoding complexity than many existing QO-STBC designs with the same code rate.

In the next section, we will also show the advantages of MDC-QOSTBC over full-diversity CIOD/ACIOD with constellation rotation, which achieve the same code rate and decoding complexity as MDC-QOSTBC.

### D. *Performance Comparison*

It has been shown in [10] that the performance of a space-time code can be optimized by maximizing the minimum determinant of the codeword distance matrix (i.e. coding gain). For practical implementation, it has further been pointed out in [8,11] that the probability, $P_o$, that an antenna transmits the "zero" symbol, should be kept as low as possible, so as to achieve a low peak-to-average power ratio. The optimum constellation rotation angle, minimum determinant (coding gain) and $P_o$ values of QO-STBC, CIOD and MDC-QOSTBC with 4QAM constellation for four transmit antennas are compared in Table 2, while their block error rates (BLER) are compared in Figure 2. These results show that our proposed MDC-QOSTBC suffers a slight 0.4 dB loss at BLER of $10^{-4}$ compared to the existing QO-STBCs (which have a higher decoding complexity), as a result of a reduced minimum determinant value. Interestingly, the same performance loss is also observed in CIOD. Hence it appears that this is a fundamental price to pay in order to achieve a lower decoding complexity. Next, comparing MDC-QOSTBC against CIOD, we observe that although they have almost identical decoding performance, our proposed MDC-QOSTBC does not require any transmit antenna to transmit zero (hence achieving the ideal value of $P_o = 0$), while CIOD requires half of the transmit antennas to transmit zero at any one time (hence $P_o = 50\%$). So our MDC-QOSTBC has an advantage over CIOD in terms of practical implementation.

Corresponding comparisons between MDC-QOSTBC, CIOD and ACIOD with 4QAM constellation for the cases of three and five transmit antennas are presented in Table 3 and Figure 3. CIOD and MDC-QOSTBC for three transmit antennas are obtained by removing the last column from their counterparts for four transmit antennas, while CIOD and MDC-QOSTBC for five transmit antennas are obtained by removing the first and last two columns from their counterparts for eight transmit antennas based on the guideline given in [8]. These results show that our proposed MDC-QOSTBC can achieve a higher minimum determinant, hence lower BLER, than CIOD. Furthermore, our code performs comparably with ACIOD and does not require any transmit antennas to transmit zero, while ACIOD for three transmit antennas requires 1/3 of the transmit antennas to be turned off at any period of time. Hence our proposed MDC-QOSTBC is more versatile in supporting both odd and even number of transmit antennas, whereas CIOD only performs well for even number of transmit antennas and ACIOD only supports odd number of transmit antennas.

## V. CONCLUSION

We have derived the generic algebraic structure of minimum-decoding-complexity Quasi-Orthogonal STBC (MDC-QOSTBC). MDC-QOSTBC has the lowest possible decoding complexity for any QO-STBC, i.e. its maximum likelihood decoding only requires a joint detection of two real symbols. A set of dispersion matrices' mapping rules is proposed to systematically construct MDC-QOSTBC for an even number of transmit antennas from O-STBCs. The optimum constellation rotation angle for the modulation to be used by MDC-QOSTBC to achieve optimum decoding performance has been found to be $13.29^0$ for square or rectangular QAM, $31.7^0$ for QPSK, and $4.9^0$ for 8PSK. Columns of an MDC-QOSTBC codeword can be truncated in order to support odd number of transmit antennas without loss of diversity gain. The maximum possible code rate for the resultant MDC-QOSTBC is shown to be 1 for three and four transmit antennas and ¾ for five to eight transmit antennas. As compared with the Co-ordinate Interleaved Orthogonal Design (CIOD) and Asymmetric CIOD (ACIOD), our proposed MDC-QOSTBC has better power distribution property as it does not require any transmit antenna to be turned off and it is more versatile in supporting different number of transmit antennas. In addition, MDC-QOSTBC has better decoding performance than CIOD for odd number of transmit antennas.

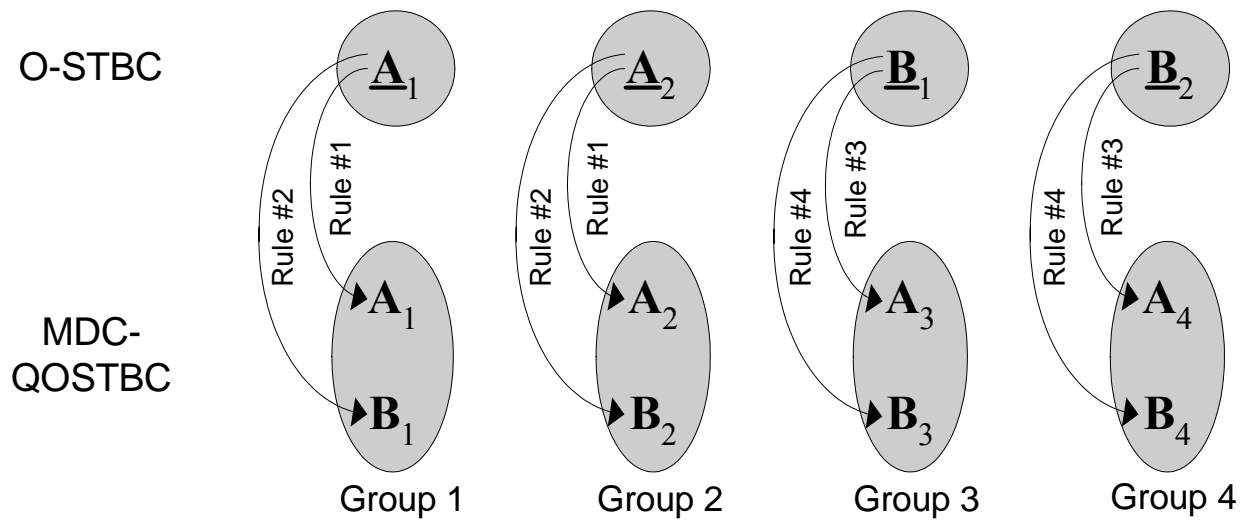

Figure 1 Construction of MDC-QOSTBC from O-STBC

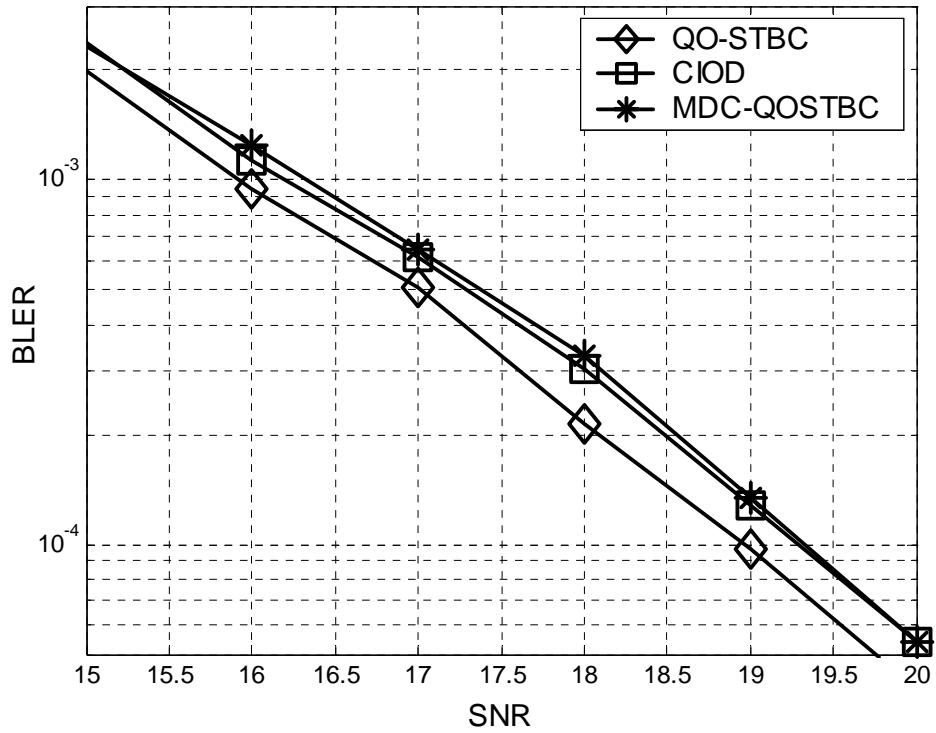

Figure 2 Simulation results for four transmit antennas with 4QAM constellation

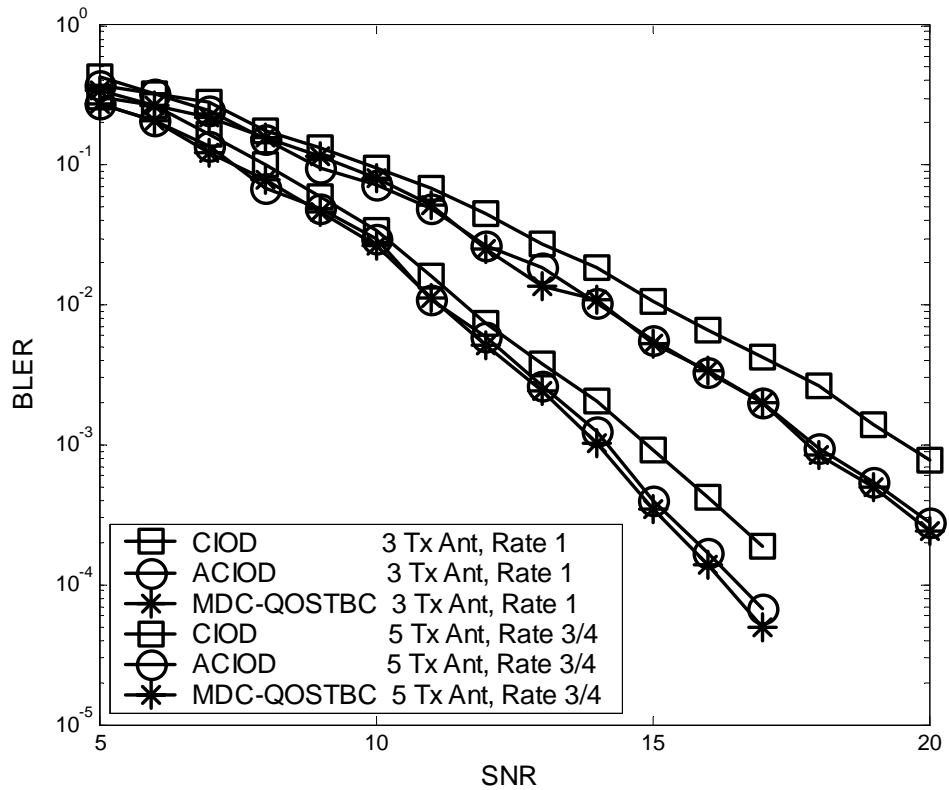

Figure 3 Simulation results for three and five transmit antennas with 4QAM constellation

Table 1 Comparison between O-STBC and QO-STBC

| Tx. Antennas | O-STBC | | | QO-STBC | | | |
|---|---|---|---|---|---|---|---|
| | Ref. | Code Length | Max. Rate | Reference | Code Length | Max. Rate | Complexity |
| 2 | [1] | 2 | 1 | N.A. | | | |
| 3-4 | [1,2] | 4 | 3/4 | [3-6] | 4 | 1 | 4 |
| | | | | CIOD/ACIOD [7,8] | 4 | 1 | 2 |
| | | | | MDC-QOSTBC | 4 | 1 | 2 |
| 5-8 | [1,2] | 8 | ½ | [3,6] | 8 | 3/4 | 4 |
| | | | | CIOD/ACIOD [7,8] | 8 | 3/4 | 2 |
| | | | | MDC-QOSTBC | 8 | 3/4 | 2 |

Table 2 Comparison of QO-STBCs for four transmit antennas

| | Optimum Constellation Angle | No. of Real Symbols for ML Joint Detection | Minimum Determinant | $P_o$ |
|---|---|---|---|---|
| QO-STBC [3] [6] | $45^0$ | 4 | 16 | 0 |
| CIOD [7] | $31.72^0$ | 2 | 10.2347 | 50% |
| MDC-QOSTBC | $13.29^0$ | 2 | 10.2347 | 0 |

Table 3 Comparison of QO-STBCs for three and five transmit antennas

| | | Three Tx Antennas | | Five Tx Antennas | |
|---|---|---|---|---|---|
| | Optimum Constellation Angle | Min Determinant | $P_o$ | Minimum Determinant | $P_o$ |
| CIOD [7] | $31.72^0$ | 0.32 | 50% | 5.56 | 50% |
| ACIOD [8] | $31.72^0$ | 5.40 | 33% | 82.28 | 20% |
| MDC-QOSTBC | $13.29^0$ | 6.40 | 0 | 107.88 | 0 |